\documentclass[pre,showpacs,twocolumn,longbibliograph,superscriptaddress]{revtex4-2}

\usepackage{hyperref}
\usepackage{color}
\usepackage[usenames,dvipsnames]{xcolor}
\usepackage{amsmath,amsthm,amssymb}
\usepackage{graphicx}
\usepackage{float}
\usepackage{epsfig}
\usepackage{bm}
\usepackage{mathrsfs}
\usepackage{multirow}
\usepackage[all]{xy}
\usepackage{pbox}
\usepackage{verbatim}
\usepackage{braket}
\usepackage{mathtools}
\usepackage{bm}
\usepackage{tikz}
\usepackage{xcolor}
 
\usepackage{mathtools}

\usepackage{mathtools}
\usepackage{enumerate}

\usepackage{acronym}

\newacro{goe}[GOE]{Gaussian orthogonal ensemble}
\newacro{P}[P]{Poisson}
\newacro{ME}[ME]{mobility edge}
\newacro{QP}[QP]{quasiparticle}
\newacro{KCM}[KCM]{kinetically constrained model}
\newacro{FOE}[FOE]{fluctuation operator expansion}
\newacro{GP}[GP]{Gross-Pitaevskii}
\newacro{EOM}[eom]{equation of motion}
\newacro{ED}[ED]{exact diagonalization}
\newacro{MF}[MF]{mean-field}

\newcommand{\beq}{\begin{equation}}
\newcommand{\eeq}{\end{equation}}

\begin{document}  

\title{Slow dynamics and non-ergodicity of the bosonic quantum East model in the semiclassical limit}

\author{Andreas Gei{\ss}ler}
\affiliation{School of Physics and Astronomy, University of Nottingham, Nottingham, NG7 2RD, UK}
\affiliation{Centre for the Mathematics and Theoretical Physics of Quantum Non-Equilibrium Systems,
University of Nottingham, Nottingham, NG7 2RD, UK}
\author{Juan P. Garrahan}
\affiliation{School of Physics and Astronomy, University of Nottingham, Nottingham, NG7 2RD, UK}
\affiliation{Centre for the Mathematics and Theoretical Physics of Quantum Non-Equilibrium Systems,
University of Nottingham, Nottingham, NG7 2RD, UK}

\begin{abstract}
We study the unitary dynamics of the bosonic quantum East model, a kinetically constrained lattice model which generalises the quantum East model to arbitrary occupation per site. We consider the semiclassical limit of large (but finite) site occupancy, so that the dynamics are approximated by an evolution equation of the Gross-Pitaevskii kind. This allows us to numerically study in detail system sizes of hundreds of sites. Like in the spin-$1/2$ case, we find two dynamical phases, an {\em active} one of fast thermalisation, and an {\em inactive} one of slow relaxation and absence of ergodicity on numerically accessible timescales. The location of this apparent ergodic to non-ergodic transition coincides with the localisation transition of the ground state. We further characterize states which are non-ergodic on all timescales in the otherwise ergodic regime.
\end{abstract}

\maketitle

\section{Introduction}

The East model is a classical and stochastic lattice spin model relevant for the study of slow dynamics as it occurs for example in supercooled liquids and glasses \cite{Jackle1991,Garrahan2002b}. It is a kinetically constrained model (KCM) \cite{Ritort2003,Garrahan2011,Garrahan2018} where transitions (spin flips) are subject to local constraints. KCMs have been shown to display very rich dynamics even when their thermodynamics is very simple. In this way, they allow to differentiate the problem of complex slow relaxation from that of static phase transitions \cite{Chandler2010}. 

The idea of slow dynamics due to kinetic constraints, rather than, say, due to quenched disorder in the energetics, is also relevant for quantum many-body systems evolving under unitary dynamics. In particular, the quantum East model has been shown to display in one parameter regime characteristics associated with non-ergodicity, such as very slow relaxation \cite{VanHorssen2015} and a proliferation of non-thermal eigenstate throughout the spectrum \cite{Pancotti2020}. 

Here we study a bosonic version of the quantum East model, also recently considered in Ref.~\cite{Valencia-Tortora2022}. We focus on the semiclassical limit, by extending the degrees of freedom at each lattice site from spin-$1/2$ to large spin, approximating them for the high-spin case by means of Holstein-Primakoff bosons \cite{Holstein1940}. As we show below, at low energies this bosonic system may form a (magnon) Bose-Einstein condensate characterized by off-diagonal long-range order, admitting a mean-field treatment with straightforward \ac{FOE} corrections \cite{Geissler2018,Geissler2019a}. The large spin limit further allows for the semi-classical treatment of the equation of motion in terms of a \ac{GP} approach \cite{Gross1961,Pitaevskii1961,Polkovnikov2002}. 
This approach allows to numerically study the dynamics of systems of hundreds of sites for very long times.

As in the spin-1/2 case \cite{VanHorssen2015,Pancotti2020}, the Hamiltonian we consider corresponds to a deformation (or ``tilting'' in the language of large deviations) of the classical stochastic generator of the bosonic East model \cite{Jack2006}. Our numerics suggest the existence of two distinct quantum phases controlled by the deformation parameter, an {\em active} phase where the ground-state is extended, and an {\em inactive} phase where the ground-state is localised (cf.\ Ref.~\cite{Valencia-Tortora2022}).
Furthermore, the Gross-Pitaevskii like dynamics shows that while the system thermalizes fast in the active phase, thermalisation is logarithmically slow in the inactive phase, with the system not being able to achieve ergodicity within numerically accessible times. Our semiclassical approch allows us to study in detail the space-time patterns of relaxation that distinguish both dynamical phases. 

The paper is organised as follows. In Sec.~II we introduce the model. In Sec.~III we consider the properties of the ground state. In Sec.~IV we consider the quantum dynamics in the semiclassical limit. In Sec.~V we give our conclusions.

\section{Model}

The classical East model is defined in terms of binary variables on a one-dimensional lattice. The {\em bosonic East model} is obtained by generalising the single occupancy of each lattice site to arbitrary occupation. The stochastic generator reads \cite{Jack2006}: 
\beq 
  \label{eq:W}
  \mathcal{W} = \sum_{\ell} n_{\ell-1} \left(
  \gamma a^{\dagger}_{\ell} + \kappa a_{\ell} - \gamma -\kappa n_{\ell}  \right)
\eeq
where the number of bosons $n_{\ell} = a^{\dagger}_{\ell} a_{\ell}$ per site $\ell$ is given in terms of their creation (annihilation) operators $a^{\dagger}_{\ell}$ ($a_{\ell}$). 
The operator \eqref{eq:W} generates dynamics where particles are created or destroyed subject to the kinetic constraint that the occupation of the nearest neighbor site to the left is non-vanishing. When the constraint is satisfied, the rate for creation, 
$n_{\ell} \rightarrow n_{\ell}+1$, is $\gamma$, 
while the rate to destroy, $n_{\ell}+1 \rightarrow n_{\ell}$, is $\kappa$. The equilibrium state of the corresponding classical stochastic dynamics is non-interacting, with site occupation independent and identically Poissonian distributed with ``rate'' $\gamma/\kappa$ \cite{Jack2006}.

The large deviations of the classical stochastic dynamics can be studied by deforming or {\em tilting} the generator \cite{Lecomte2007,Touchette2009,Garrahan2018, Jack2020}. For the statistics of the dynamical activity (i.e., the number of configuration changes in a trajectory \cite{Lecomte2007,Garrahan2007,Maes2020}) the tilted generator reads,
\beq 
  \label{eq:Ws}
  \mathcal{W}_s = \sum_{\ell} n_{\ell-1} \left[
   e^{-s} \left(
  \gamma a^{\dagger}_{\ell} + \kappa a_{\ell} 
  \right)
  - \gamma -\kappa n_{\ell}  \right]
\eeq
where the parameter $s$ controls the tilting. 

The generator ${\cal W}$ obeys detailed balance, and so does the titled one ${\cal W}_s$. This means that we can make ${\cal W}_s$ Hermitian via a similarity transformation using the equilibrium state of ${\cal W}$, which we know as it is a simple product state, see e.g.~\cite{Jack2006}. 
Specifically, if we make the transformations
$
a^{\dagger}_i \rightarrow \hat{b}^{\dagger}_i \sqrt{\frac{\kappa}{\gamma}}
$
 and 
$
a_i \rightarrow \sqrt{\frac{\gamma}{\kappa}} \hat{b}_i
$, 
the corresponding Hamiltonian takes the form
(we set $\hbar=1$ throughout)
\beq
\label{Hs}
H_s = - \sum_{\ell} \hat{n}_{\ell-1} \left[ e^{-s} \sqrt{\kappa \gamma} \left( \hat{b}^{\dagger}_{\ell} + \hat{b}_\ell \right) - \gamma -\kappa \hat{n}_\ell  \right].
\eeq

For $s<0$ the Hamiltonian above will lead to runaway occupation. To prevent this from happening in our numerics below we will truncate the occupation numbers using Holstein-Primakoff bosons (see also how this issue is alteratively addressed in Ref.~\cite{Valencia-Tortora2022}). 
Without loss of generality we set $\kappa = 1$ and $\gamma = (N-\hat{n}_\ell)\tilde{c}$ where $N$ is the maximal number of bosons per site and $\tilde{c} = c/N$. The Hamiltonian thus reads $H = - \sum_{\ell} \mathcal{C}_{\ell} \hat{h}_{\ell}$ in terms of the local generator
\beq \label{eq:loc_generator_Herm}
\hat{h}_{\ell}(s) = e^{-s} \sqrt{\tilde{c}} \left( \hat{b}^{\dagger}_{\ell} \sqrt{N-\hat{n}_\ell} + \textrm{h.c.} \right) - \tilde{c}(N-\hat{n}_\ell) - \hat{n}_\ell,
\eeq
which implies the possibility of product eigenstates. Besides the trivial zero energy vacuum state, for $s=0$ each site may also be the lowest eigenstate of $-\hat{h}_{\ell}(0) | \alpha \rangle = 0 | \alpha \rangle$, 
which is of binomial form in the Fock-states $| n \rangle$:
\begin{align} 
| \alpha \rangle &= \sum_{n=0}^N \frac{\alpha^{\frac{n}{2}}}{\left(1+|\alpha| \right)^{\frac{N}{2}}} \sqrt{\binom{N}{n}} | n \rangle \textrm{ with} \\
\label{eq:binom_state_ocp}
\langle \alpha | \hat{n} | \alpha \rangle &= \frac{N\alpha}{1+|\alpha|} \textrm{ and }
\langle \alpha |\sqrt{N-\hat{n}} \hat{b} | \alpha \rangle = \frac{N\sqrt{\alpha}}{1+|\alpha|}.
\end{align} 
with $\alpha = \tilde{c}$. For $N \rightarrow \infty$ expectation values tend towards their coherent values, so $\langle \alpha | \hat{n} | \alpha \rangle |_{N \rightarrow \infty} = c$. 
As the product \ac{MF} state $| \psi \rangle = \prod_{\ell}|\alpha\rangle_{\ell}$ is an eigenstate for $s=0$, independent of the boundary conditions, the number of spatial dimensions $d$ or the kinetic constraint, we expect the \ac{MF} ansatz to be a good approximation also for $s \neq 0$ even for $d=1$ (assuming continuity).

\begin{figure*}[t]
  \centering
  \includegraphics[width=0.98\textwidth]{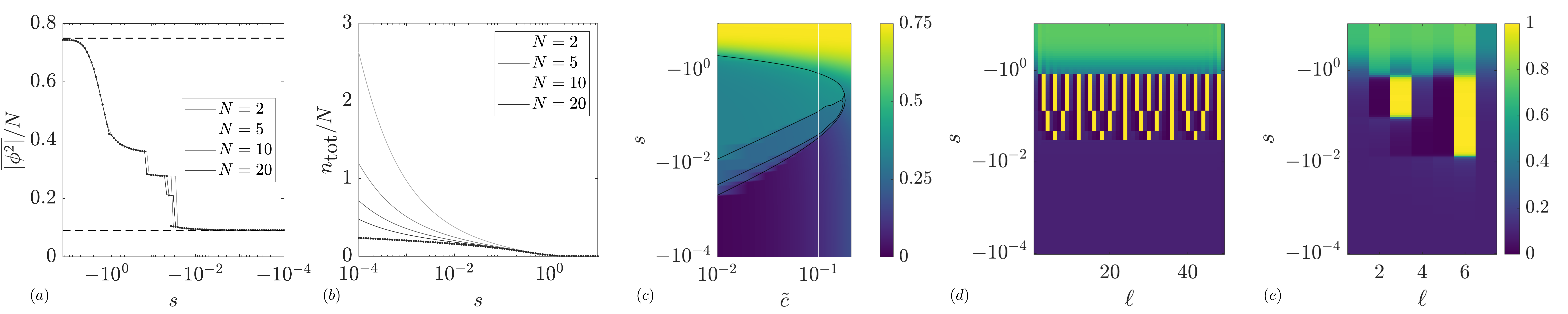}
  \caption{{\bf Ground state.}
 (a) Filling fraction $\overline{|\phi^2|} = \langle \psi | \sum_{\ell} \hat{n}_{\ell} | \psi \rangle / L$ for $s<0$, scaled by truncation parameter $N$, at $\tilde{c}=0.1$. Results are from \ac{MF} (dots) and \ac{FOE} (solid lines). Dashed lines refer to limiting values discussed in the main text.
 (b) Same but for $s>0$, where $n_{\textrm{tot}} = L \overline{|\phi^2|}$. 
 (c) Filling $\overline{|\phi^2|}$ in the Gross-Pitaevskii limit as function of $s$ and $\tilde{c}$. Regions delimited by black lines correspond to unit cell sizes. The white vertical line indicates $\tilde{c}=0.1$. (d) Density profiles from \ac{FOE} with $N=20$ and $L=49$ for $s<0$. (e) Same from \ac{ED} with $N=3$ and $L=7$.
 }
  \label{fig:groundstate}
 \end{figure*}

\section{Localisation transition in the ground state}

We determine the ground state of the Hamiltonian \eqref{Hs} by comparing three numerical methods.

\subsection{Mean field}

The first of these corresponds to \ac{MF} supplemented by \ac{FOE} quasiparticle corrections. We start by minimizing the energy of a product \ac{MF} ansatz, $| \psi \rangle = \otimes_{\ell=1}^L | \phi_\ell \rangle$. 
We define the fluctuation operators $\delta \hat{n}_{\ell} \equiv \hat{n}_{\ell} - n_{\ell} $ and $\delta \hat{h}_{\ell} \equiv \hat{h}_{\ell} - h_{\ell}$, with real valued shifts $n_{\ell}$ and $h_{\ell}$. The Hamiltonian takes the form 
\begin{align}
  H_s = \sum_{\ell=1}^L n_{\ell-1} \hat{h}_{\ell} + \hat{n}_{\ell} h_{\ell+1} - n_{\ell}h_{\ell+1} + \delta \hat{n}_{\ell} \delta \hat{h}_{\ell+1}  
\end{align}
Discarding the terms of second-order in the fluctuation terms $\delta \hat{n}_{\ell}$ and $\delta \hat{h}_{\ell}$ yields the standard \ac{MF} decoupling in terms of a sum over the local contributions, $-n_{\ell-1} \hat{h}_{\ell} - \hat{n}_{\ell} h_{\ell+1} + n_{\ell}h_{\ell+1}$. 
All of these have as their ground state $| 0 \rangle_{\ell}=| \alpha_{\ell} \rangle$ of the form \eqref{eq:binom_state_ocp}. This combines into the global \ac{MF} ground state $| \psi_{\textrm{MF}} \rangle = \otimes_{\ell} | 0 \rangle_{\ell}$. For each term the self-consistency conditions $n_{\ell} = {}_{\ell} \langle 0 | \hat{n}_{\ell} | 0 \rangle_{\ell} $ and $h_{\ell} = {}_{\ell} \langle 0 | \hat{h}_{\ell}| 0 \rangle_{\ell}$ have to be fulfilled and $\alpha = (x+\sqrt{x^2+y^2})^2/y^2$ with $x=[h_{\ell+1}/n_{\ell-1}-(1-\tilde{c})]$ and $y=2\sqrt{\tilde{c}}\exp(-s)$.

\subsection{Fluctuation operator expansion}

We obtain corrections to the \ac{MF} result via the \ac{FOE} method. To do so the eigenbasis $\lbrace | k \rangle_{\ell} \rbrace$ with energies $E_k^{(\ell)}$ ($k \in [0,N]$) for each $H_{\ell}$ is used to expand the 
$\delta \hat{n}_{\ell} \delta \hat{h}_{\ell+1}$ terms using local raising (lowering) operators $\sigma^{\dagger}_{k,\ell} = | k \rangle_{\ell} {}_{\ell} \langle 0 |$ ($\sigma_{k,\ell} = | 0 \rangle_{\ell} {}_{\ell} \langle k |$). As discussed elsewhere \cite{Geissler2018,Geissler2019a,Geissler2020}, this expansion yields an exact representation of the Hamiltonian with terms up to fourth order in the $\sigma$-operators which generate the dynamics of local fluctuations and their interactions. Using a generalized Bogoliubov transformation of the form $\beta_{\alpha} = \sum_{k,\ell} u^{* \alpha}_{k,\ell} \sigma_{k,\ell} + v^{* \alpha}_{k,\ell} \sigma^{\dagger}_{k,\ell}$ with the normalization $\sum_{k,\ell}u^{* \alpha}_{k,\ell}u^{\alpha}_{k,\ell} - v^{* \alpha}_{k,\ell}v^{\alpha}_{k,\ell} = 1$ one finds the form
\begin{align}
  \label{eq:H_FOE_form}
  H = \sum_{\ell} E_0^{(\ell)} + \frac{1}{2}\left( \sum_{\alpha} \omega_{\alpha}  - \textrm{Tr}[h] \right) 
  \\
  \nonumber
  + \sum_{\alpha} \omega_{\alpha} \beta^{\dagger}_{\alpha} \beta_{\alpha} + \mathcal{H}
\end{align}
where $\omega_{\alpha}$ are the positive eigenvalues of 
\begin{align}
  H_{\textrm{FOE}} = 
  \begin{pmatrix}
  h & \Delta \\
  -\Delta* & -h* 
  \end{pmatrix}    
\end{align}
given in terms of the matrix elements 
\begin{align}
  h_{i\ell,j\ell'} = (E_i^{(\ell)} & - E_0^{(\ell)}) \delta_{i,j}\delta_{\ell,\ell'} 
  \\
  \nonumber  
  &
  - {}_{\ell}\langle i| \delta\hat{n}_{\ell} | 0 \rangle_{\ell} {}_{\ell'}\langle 0| \delta\hat{h}_{\ell'} | j \rangle_{\ell'} \delta_{\ell',\ell+1}
  \end{align}
and 
\begin{align}
  \Delta_{i\ell,j\ell'} = - {}_{\ell}\langle i| \delta\hat{n}_{\ell} | 0 \rangle_{\ell} {}_{\ell'}\langle j| \delta\hat{h}_{\ell'} | 0 \rangle_{\ell'} \delta_{\ell',\ell+1}
\end{align}
The first and second term of~\eqref{eq:H_FOE_form} are the \ac{MF} energy and a scalar correction while the third and fourth terms are the non-interacting quasiparticle modes and their interactions. With all $\omega_{\alpha} > 0$ we proceed to define the quasiparticle ground state implicitly by demanding $\beta_{\alpha} | \psi_{\textrm{FOE}} \rangle= 0$ for all $\alpha$. The \ac{FOE} ground state is then determined by minimizing $\langle \psi_{\textrm{FOE}} | \mathcal{H} | \psi_{\textrm{FOE}} \rangle$ over a  set of self-consistent \ac{MF} states consisting of ones with variable high-density peak spacing $l \in \left[ 1,2,3,4,6,12 \right]$ as well as one without high density peaks ($\langle \phi | \hat{n} | \phi \rangle < 3L/4$ for all sites).

\subsection{Results}

The results of the MF and the FOE are shown in Figs.~\ref{fig:groundstate}. While there are no appreciable finite-size effects in either in terms of $L$, the scaling with respect to the truncation $N$ is shown for in terms of the filling fraction $\overline{|\phi^2|}/N = \langle \psi | \sum_{\ell=1}^L \hat{n}_{\ell} | \psi \rangle / LN$ $s<0$, Fig~\ref{fig:groundstate}(a), and for the total particle number $n_{\textrm{tot}} = L \overline{|\phi^2|}$ for  $s>0$, Fig~\ref{fig:groundstate}(b). For $s<0$ the filling fraction is bounded between the exact \ac{MF} result $\tilde{c}/(1+\tilde{c})$ for $s=0$ and $3/4$ for $s \rightarrow - \infty$. The visible steps indicate density-wave patterns, see 
Figs.~\ref{fig:groundstate}(d,e) for $\tilde{c}=0.1$. 

While the \ac{MF} result does not have a significant dependence on $N$ for $N>2$ for any value of $s$, the 
\ac{FOE} approximation in contrast does. For $s>0$ the system is inactive and 
only sites close to the boundary ($n_0 = N$) are appreciably occupied. Therefore we consider the total filling instead. Both, the \ac{MF} and the \ac{FOE} total filling, diverge for $s \rightarrow 0$, consistent with a homogeneous \ac{MF} state at $s=0$. While the \ac{FOE} divergence is stronger, it tends towards \ac{MF} behavior for increasing $N$.

For the $N \rightarrow \infty$ we further consider the \ac{GP}-like limit for which the ground state minimizes the energy~\eqref{eq:GP_energy}, see discussion below. As shown in Fig.~\ref{fig:groundstate}(c), this yields the same density-wave steps as in the \ac{MF}, \ac{FOE}, cf Fig.~\ref{fig:groundstate}(d). For comparison, in Fig.~\ref{fig:groundstate}(e) we show the corresponding results from \ac{ED}. Within the \ac{GP} limit we find a critical $\tilde{c} \approx 0.18$ above which no density-wave pattern can be observed in the ground state.

\bigskip

\begin{figure*}[ht!]
  \centering
  \includegraphics[width=0.98\textwidth]{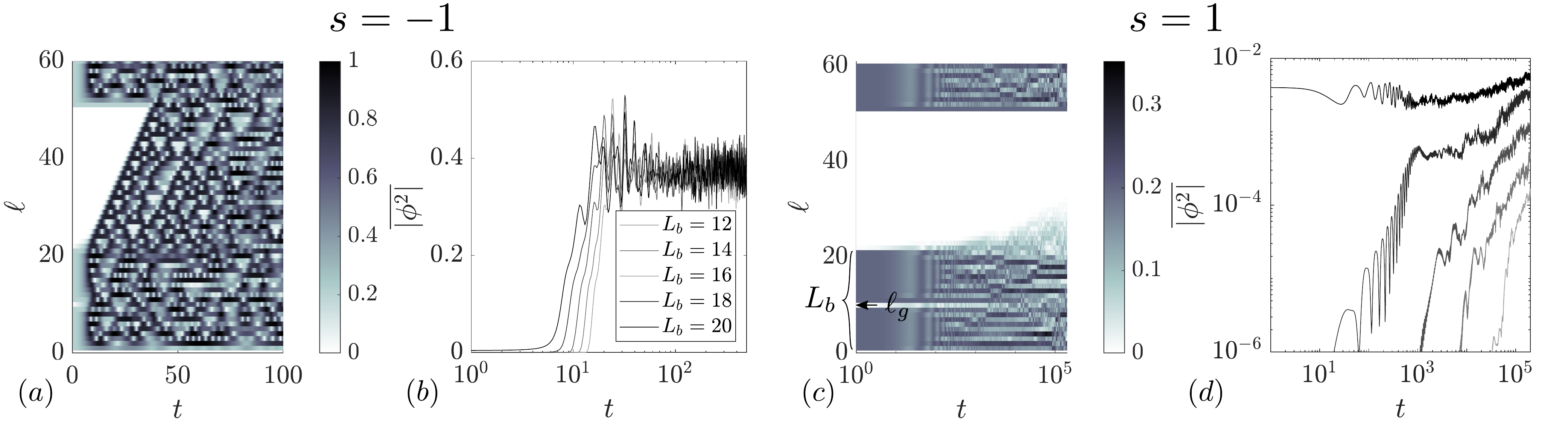}
  \caption{
    {\bf Dynamics after a quench.}
   Trajectories of~\eqref{eq:GP_EAST_dyn} in the active ($s=-1$ in $a,b$) and inactive ($s=1$ in $c,d$) cases. While $(a,c)$ show sample trajectories of $\phi_{\ell}$ with $L_b=20$ and $\ell_g=10$ [marked in (c)], sub-system densities $\overline{|\phi^2|} = \sum_{\ell=21}^{30} |\phi^2_{\ell}|/10$ sampled over $\ell_b \in [1,10]$ are shown in $(b,d)$ for various widths $L_b$ (see legend). In the active case the densities thermalize within $t<100$ for any $L_b$ of initially activated (single-gapped-)block while no thermal or steady-state is reached for times $t=2\cdot 10^5$ and $s=1$. 
  }
  \label{fig:GPdyn_locVSdeloc}
 \end{figure*}

\section{Dynamics in the semiclassical limit}

Considering the mean-field like behavior of the ground states as $N$ is increased, we take the semiclassical limit $N \rightarrow \infty$ to study dynamics in terms of a \ac{GP}-like equation. We obtain this via the standard replacement $\hat{b}_{\ell} \rightarrow \sqrt{N} \phi_{\ell}$. In this limit the homogeneous ground state for $s=0$ has $|\phi_{\ell}^2| = |\phi^2| = \tilde{c}/(1+\tilde{c})$, cf.\ Eq.~\eqref{eq:binom_state_ocp}. Furthermore, the energy functional has to be regularized as
\begin{align} \label{eq:GP_energy}
  \mathcal{E} =& \frac{E}{N^2} = \sum_{\ell=1}^{L} \mathcal{E}_{\ell} \equiv \sum_{\ell=1}^{L} |\phi_{\ell-1}|^2 h(\phi_{\ell})
\end{align}
with
\begin{align}
  h(\phi) =& (1-\tilde{c}) |\phi|^2 + \tilde{c} - e^{-s}\sqrt{\tilde{c}} (\phi + \phi^*) \sqrt{1-|\phi|^2}.
\end{align}c
From the Heisenberg equation of motion
we derive the \ac{GP}-like equation of motion with rescaled time $t \rightarrow tN$ in the limit $N \rightarrow \infty$:
\begin{align}
  \dot{\phi}_{\ell} = & \frac{\phi_{\ell}}{i\hbar} \left[ \tilde{c} + (1-\tilde{c})|\phi_{\ell+1}|^2 - e^{-s}\sqrt{\tilde{c}}\sqrt{1-|\phi_{\ell+1}|^2} 2 \Re(\phi_{\ell}) \right] 
  \nonumber
  \\ &
  + \frac{|\phi_{\ell-1}|^2}{i\hbar} 
  \left[   
    (1-\tilde{c})\phi_{\ell} - e^{-s} \sqrt{\tilde{c}} 
    \phantom{
      \left( 
        \sqrt{1-|\phi_{\ell}|^2}
        -  
        \frac{\phi_{\ell} \Re(\phi_{\ell})}{\sqrt{1-|\phi_{\ell}|^2}} 
      \right)
    }  
  \right.
  \nonumber
  \\
  &
  \left.
    \phantom{
      (1-\tilde{c})\phi_{\ell} 
    }
    \times
    \left( 
      \sqrt{1-|\phi_{\ell}|^2}
      -  
      \frac{\phi_{\ell} \Re(\phi_{\ell})}{\sqrt{1-|\phi_{\ell}|^2}} 
    \right)
  \right]
\end{align} \label{eq:GP_EAST_dyn}
The final singularity limits the complex field $\phi_{\ell}$ to stay within the physically meaningful disc $|\phi_{\ell}|^2 < 1$.
Depending on boundary and initial conditions one can identify distinct dynamical regimes.

\subsection{Relaxation dynamics after a quench: active to inactive transition}

From here on we consider fixed boundary conditions of the form $\phi_0 = 0.2$ and $\phi_{L+1}= 0$ unless specified otherwise. With these, the simple dynamics above quickly gives way to chaotic non-linear dynamics. The general picture is summarized in Fig.~\ref{fig:GPdyn_locVSdeloc} for $\tilde{c} = 0.1$. There, we set $L=60$ and always start with $L_i = 30$ sites initially activated to $\phi_0$. By keeping the number of gaps between active sites fixed we can guarantee identical initial energies. As sample cases we split the set of initially active sites into a block of $L_b+1$ sites starting at $\ell = 1$ and the remaining $L_i-L_b$ at the end of the system. We further want to sample over similar initial conditions by adding a gap at site $\ell_g$ in the first block. To determine whether the system thermalizes we look at the sub-system densities $\overline{|\phi^2|} = \sum_{\ell=21}^{30} |\phi^2_{\ell}|/10$ sampled over $\ell_g \in [1,10]$. 

In the active phase $s<0$ (we show the example of $s=-1$ in Fig.~\ref{fig:GPdyn_locVSdeloc}) we observe a ballistic spreading of active sites over the entire system. Note in particular how the the initially empty sites only reach $|\phi_{\ell}| \approx 1$ after an extended time interval compared to the rest of the system. This is an evident consequence of the kinetic constraints that force empty regions to relax from their boundaries; we discuss this pre-thermal dynamics in further detail below. Independently of the width $L_b$ of the first block, the system always reaches its thermal state at $t\approx 100$, see Fig.~\ref{fig:GPdyn_locVSdeloc}(b). Differences at early times are related to the gap between the first block and the observed sub-system and reflect the ballistic spreading of excitations. 

The behaviour in the inactive phase $s>0$ is very different (see $s=1$ in Fig.~\ref{fig:GPdyn_locVSdeloc}). Here excitations only spread logarithmically in time from the initially active into the empty region, reminiscent of what occurs in the spin-1/2 quantum East model \cite{VanHorssen2015}. 
From the sub-system densities we see that the system does not thermalize for times up to at least $t=10^5$ while the values for different block sizes $L_b$ always remain distinct within the observed time window.

\begin{figure*}[ht!]
 \centering
 \includegraphics[width=0.98\textwidth]{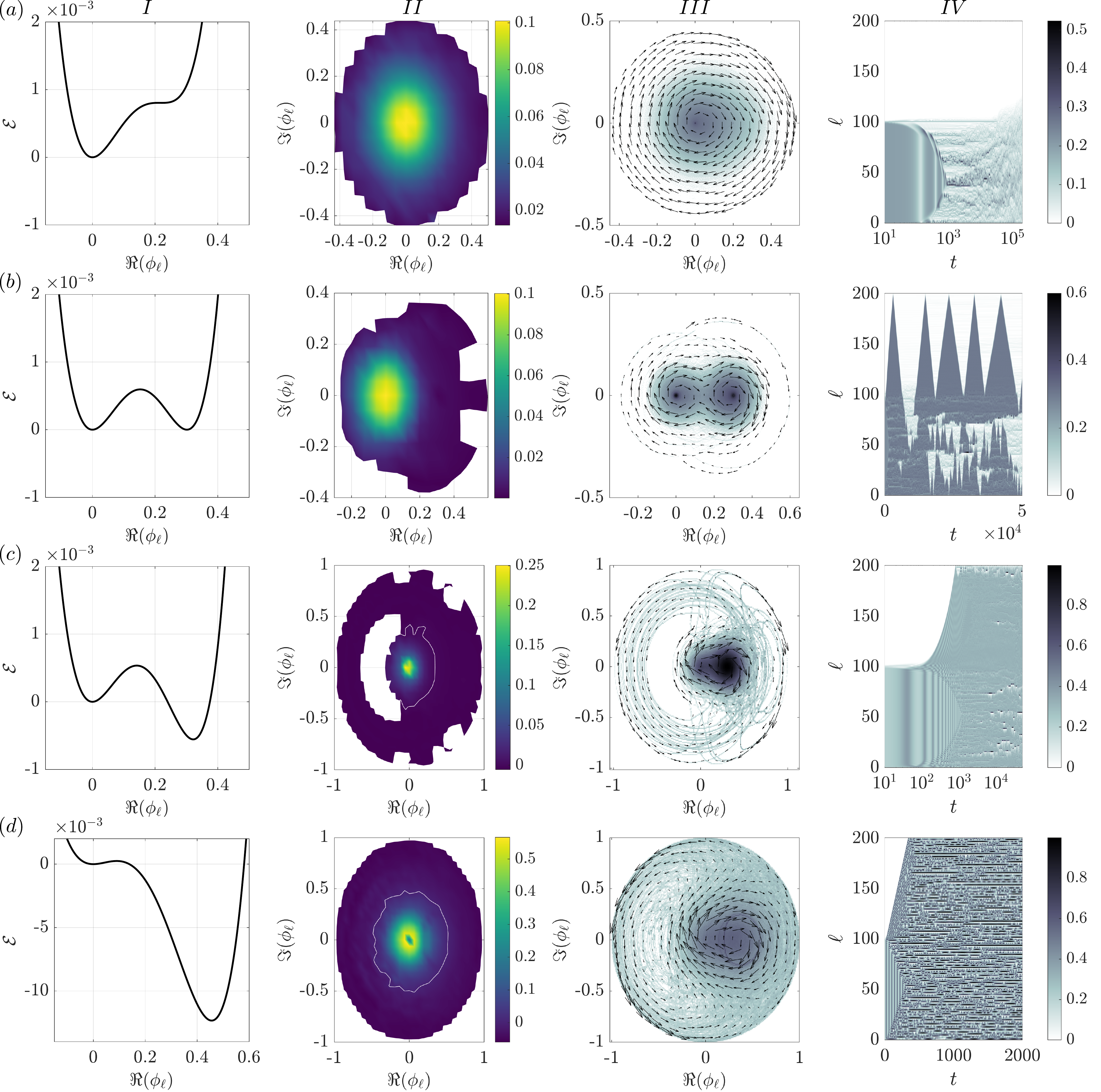}
 \caption{
  {\bf Phase portrait of dynamics.}
Overview of typical dynamical regimes in the \ac{GP}-like time evolution~\eqref{eq:GP_EAST_dyn} for $\tilde{c}=0.1$ and $s \in [0.067,0,-0.03,-0.3]$ respectively corresponding to rows $(a-d)$. Each case is characterized by a distinct homogeneous energy profile $\mathcal{E}$ $I$, time-and-space averaged local generator $h_{\textrm{EAST}}$ to the east $II$ with the white line marking its sign change, time-and-space sampled probability distribution of $\phi_{\ell}$ with black arrows indicating the average flow rate $\overline{\dot{\phi}}$ $III$ and the corresponding trajectories $|\phi_{\ell}|$ $IV$ for times up to $2\cdot10^5$.
 }
 \label{fig:GPdyn_atlas}
\end{figure*}

\subsection{Phase space description of active-inactive transition}

In the simplest case of a homogeneous initial state $\phi_{\ell} = \phi$ the dynamics
due to Eq.~\ref{eq:GP_EAST_dyn} is determined by a fixed total energy 
\begin{align}
  \mathcal{E}(\phi)/L = |\phi|^2 \left[ (1-\tilde{c}) |\phi|^2 + \tilde{c} - 2 e^{-s}\sqrt{\tilde{c}} \sqrt{1-|\phi|^2}\Re(\phi) \right]
  \nonumber
\end{align}
In this case, and for periodic boundaries, $\phi$ oscillates along closed loops of fixed energy $\mathcal{E}(\phi)/L$. These collective oscillations cab be even visible for initially homogeneous sub-systems as in Fig.~\ref{fig:GPdyn_locVSdeloc}. 

The dynamical phases of the bosonic quantum East, at the level of this semiclassical description, can be understood by inspecting $\mathcal{E}$. In general, this homogeneous energy may have two minima, one at $\phi=0$ from the $|\phi|^2$ factor with energy equal to zero, and a second local minimum at $\phi>0$, with a saddle point in between. The former is due to the constraint and its value is always zero, 
while the latter is the minimum of the local unconstrained (tilted) generator in this representation, and its value is determined by 
$\tilde{c}$ and $s$. This suggests the possibility of a first-order transition when the global minimum of $\mathcal{E}$ changes between that due to the constraint and that due to the unconstrained generator. This is shown in column I of Fig.~\ref{fig:GPdyn_atlas}where we plot $\mathcal{E}$ vs $\Re(\phi)$ with $\Im(\phi)=0$, at fixed $c$ and for decreasing values of $s$: for $s>0$, the global minium is at $\Re(\phi)=0$, indicative of the inactive dynamics, while for $s<0$ the global minimum at a $\Re(\phi) > 0$, indicative of active dynamics. Coexistence occurs at the stochastic point, $s=0$. 

To characterize the dynamics in terms of the structure of $\mathcal{E}$ we analyze trajectories in the following manner.  
We consider a system of $L=200$ at $\tilde{c}=0.1$, where the first $L=100$ sites are initialized in an active state, $\phi_{\ell}(t=0) = 0.2$. In column IV of Fig.~\ref{fig:GPdyn_atlas} we show the  trajectories corresponding to the same values of $s$ as in column I. 
From these trajectories we first compute the local generator $h(\phi)$ next to each site (in the direction of the constraint) for every value of $\phi$ taken along the trajectory. This is shown in column II of Fig.~\ref{fig:GPdyn_atlas}. From the same trajectories we also compute the probability distribution $P(\phi)$ of $\phi_{\ell}$ together with the average velocity $\dot{\phi}(\phi)$. The corresponding phase portraits are shown in column III, showing both the density and the average velocity field.  

The different dynamical regimes are closely related to the properties of the homogeneous energy profile. Deep in the inactive phase $s>0$, Fig.~\ref{fig:GPdyn_atlas}(a), there is only the minimum at $\phi=0$, with no signature of bistable dynamics: 
the local generator is always positive (column II), the phase portrait of the dynamics (column III) shows motion around the origin, while the trajectory (column IV) shows that activity fails to propagate to the initially empty half of the system, with only late slow logarithmic spreading. 

At coexistence $s=0$, Fig.~\ref{fig:GPdyn_atlas}(b), since the initial values of the field in the active half of the system are near the unstable saddle point of $\mathcal{E}$, we see bistable dynamics: the phase portrait (column III) shows an alternation between motion around the origin and motion away from it, and since the local generator is always positive (column II), highly-activated detours of the field are rare requiring precedent sites to become close to empty. Correspondingly, the trajectory (column IV) 
shows ballistic excitation/de-excitation of the initially inactive half of the system. 

For $s=-0.03$, Fig.~\ref{fig:GPdyn_atlas}(c), the active state is the stable one, while the inactive one is only metastable, according to the homogeneous energy $\mathcal{E}$. In this case, after a slow transient, initially empty sites are activated following a ballistic spread of the active region, as seen in the trajectory (column IV). The local generator (column II) is no longer positive everywhere, while the metastability of the inactive state is evident in the phase portrait (column III) where sites remain mostly cycling the global minimum associated with a negative generator. In addition, large $|\phi_{\ell}|$ detours may occur as the system begins to thermalize for $t \gtrsim 10^4$, requiring only a few preceding (nearly) empty sites. Eastward neighbors of highly activated sites are usually in the vicinity of the local generator minimum resulting in a sign change of $h(\phi)$ for sufficiently large amplitudes of $\phi$. Consequently, $\dot{\phi}(\phi)$ can point counterclockwise whenever the first term of~\eqref{eq:GP_EAST_dyn} dominates. 

Finally, for $s=-0.3$, Fig.~\ref{fig:GPdyn_atlas}(d), the system is 
deep in the active phase with fast ballistic spreading from the initial active region into the initially inactive region, as shown in the trajectory (column IV). 
Both, the homogeneous initial dynamics ($\ell=1\ldots 100$) and the pre-thermal dynamics ($\ell=101\ldots 200$), are visible as cycles in the phase portrait (column III). At long times the trajectories cover the whole disc of $\phi$, indicative of thermalization. Counter propagating large amplitude detours of $\phi_{\ell}$ enabled by negative $h$ (column II) are entirely common and accompanied by a single preceding (nearly) empty site.

\subsection{Non-ergodicity of the $h(\phi)=0$ manifold}

In Figs.~\ref{fig:GPdyn_locVSdeloc}(a) and \ref{fig:GPdyn_atlas}(d) one can observe pre-thermal dynamics for initially empty sites where values of $|\phi_\ell|\approx 1$ are only reached at later times. These special dynamics are related to the occurrence of an extended zero of $h(\phi)$ for any $s<0$. Any site with a $\phi_{\ell}$ on (or close to) the $h(\phi_{\ell})=0$ manifold suppresses the first term of eq.~\eqref{eq:GP_EAST_dyn} in favor of the second term. The remaining dynamics are (almost) identical to the dynamics of the local generator modified by the density $|\phi_{\ell-1}|^2$ of the previous site. Dynamics on this limit cycle are quasi-stable on some site $\ell$ insofar as any (perpendicular) fluctuation in $\phi_{\ell}$ drives the site $\ell - 1$ away from the manifold, resulting in the eventual demise of these transient non-ergodic or pre-thermal dynamics.

\begin{figure}[h]
 \centering
 \includegraphics[width=0.95\columnwidth]{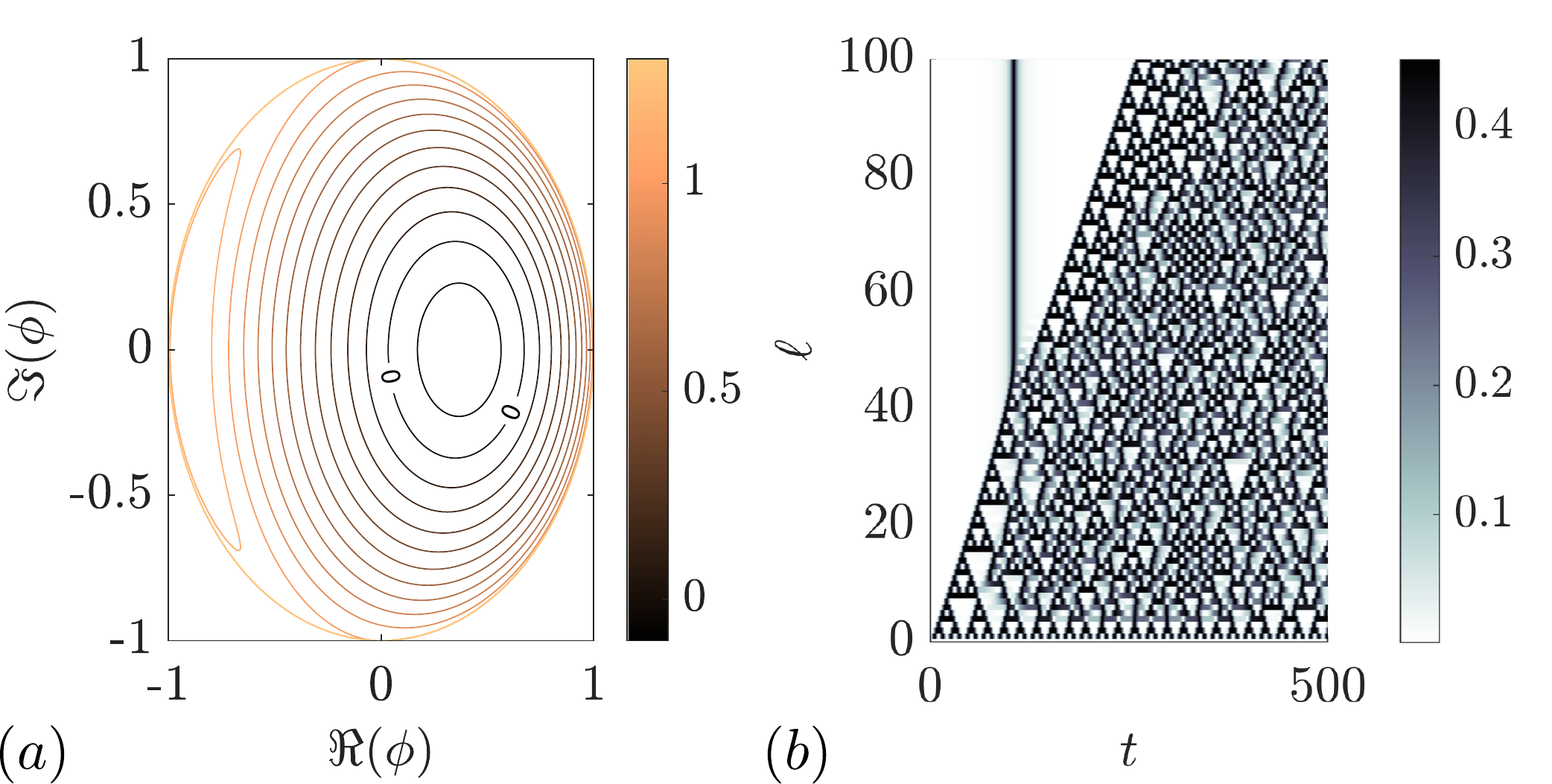}
 \caption{
  {\bf Non-ergodic dynamics on the $h(\phi)=0$ manifold.}
(a) Contour lines of $h(\phi)$ for $s=-1$ and $\tilde{c}=0.03$. The zero contour is indicated by the labels. (b) The trajectory of $|\phi_{\ell}|$ for the initial condition $\phi_0=\phi_-$ and $\phi_{\ell} = \phi_+$ for all $\ell \in [1,L+1]$ with $L=100$.
 }
 \label{fig:GP_nonergodic}
\end{figure}

An entire system on the $h(\phi)=0$ manifold will thus remain there for all times. For given $s<0$ and $\tilde{c}$ this manifold (zero contour in Fig.~\ref{fig:GP_nonergodic}) is given by 
\begin{align}
  \Im(\phi)^2 = \frac{ \left[2(1-\tilde{c})^2+4e^{-2s}\tilde{c}\right] \Re(\phi)^2+2\tilde{c}(1-\tilde{c})}{-2(1-\tilde{c})^2} 
  \\
  + \frac{2|\Re(\phi)|e^{-s}\sqrt{\tilde{c}}}
  {(1-\tilde{c})^2}
  \sqrt{e^{-2s}\tilde{c}\Re(\phi)^2+1-\tilde{c}}
  \nonumber
\end{align} \label{eq:GP_loc_gen_zero}
where $\Re(\phi)$ is bounded between 
\begin{equation}
\begin{aligned}
\phi_{\pm} = \sqrt{ \frac{\tilde{c}^2+(2e^{-2s}-1)\tilde{c} \pm 2\tilde{c}e^{-s}\sqrt{e^{-2s}-1}}{\tilde{c}^2+2(2e^{-2s}-1)\tilde{c}+1} }
\end{aligned} \label{eq:GP_loc_gen_zero_limits}
\end{equation}

Initializing all sites on the minimal value $\phi_-$ we can suppress full thermalization (see Fig.~\ref{fig:GP_nonergodic}) indicating the possibility of non-ergodic dynamics even on the typically ergodic side $s<0$. 
Notably, the occupation trajectory of this dynamics has the appearance of a random triangular fractal, especially at early times for a homogeneous initial state. This is also reminiscent of the trajectories of the classical East model \cite{Garrahan2002}.

\section{Conclusions}

In this work we studied the dynamical properties of the bosonic quantum East model, a generalisation of the spin-$1/2$ quantum East model. We found that the bosonic East model has two dynamical phases separated by a quantum-first order transition. In one phase, which we call active, the ground state is extended and thermalisation is fast. The second phase, which we call inactive, the ground state is localised and thermalisation is slow. This is analogous to what one sees in the spin-$1/2$ case \cite{VanHorssen2015,Pancotti2020}. We studied the dynamics in the semiclassical limit within a Gross-Pitaevskii approximation, which allowed us to access large system sizes in our numerics. 

Our results are closely related to those of Ref.~\cite{Valencia-Tortora2022}. In that work also a version of the bosonic quantum East model was studied in detail, finding a family of many-body localised states which allowed to create composite excitations with long-time memory. One result of Ref.~\cite{Valencia-Tortora2022} is that the bosonic East model has a localisation transition in its ground state only in the presence of density-density interactions. While we consider a slighlty different model as we restrict the site occupation to a large but finite value, 
we find something similar in our case: the diagonal interaction terms in Eq.~\eqref{Hs} appear naturally from the escape rate (the part coming from particle annihilation) of the corresponding stochastic generator Eq.~\eqref{eq:Ws}. The absence of localisation without those terms can be understood from the companion classical stochastic problem: setting them to zero is equivalent to studying rare trajectories with a very large escape rate for annihilation events, which naturally would put the system in an active (and thus extended) phase. 
Our results here, together with those of Ref.~\cite{Valencia-Tortora2022}, suggest that other quantum KCMs with interesting dynamics, such as the PXP model \cite{Fendley2004,Lesanovsky2011,Bernien2017,Turner2018} or the quantum Fredrickson-Andersen model \cite{Hickey2016}, will be worth studying in their bosonic versions, in particular in the semiclassical limit.

\bigskip

\begin{acknowledgments}
AG gratefully acknowledges support from the Leopoldina Fellowship Programme of the German National Academy of Sciences Leopoldina grant no.\ LPDS 2018-14. 
JPG acknowledges financial support from EPSRC Grants EP/R04421X/1 and EP/P034616/1. 
We acknowledge access to the University of Nottingham Augusta HPC service. 
\end{acknowledgments}

\bibliography{bosonicEAST}
\bibliographystyle{apsrev4-1}

\end{document}